\documentclass[12pt,aps,amssymb,amsmath,showpacs,prb]{revtex4}
\usepackage{graphicx}
\usepackage{dcolumn}
\usepackage{bm}
\usepackage{natbib}

\begin{document}

\title{The one-particle Green's function of one-dimensional insulating materials}
\author{M.C. Refolio}
\email{refolio@imaff.cfmac.csic.es}
\author{ J.M. L\`{o}pez Sancho}
\author{J. Rubio}

\affiliation{Instituto de Matem\`{a}ticas y F\`{\i}sica
Fundamental, CSIC\\
Serrano 113 bis, E28006 Madrid, Spain}
\author{J.A. Verges}
\affiliation{Instituto de Ciencia de Materiales ICMM-CSIC,
\\
Cantoblanco, E28049 Madrid, Spain}
\date{\today}

\begin{abstract}
The single particle spectral-weight function (SWF) of the ionic
Hubbard model at half filling is calculated in the cluster
perturbation theory approximation. An abrupt change of regime in
the low-energy region, near the chemical potential, is found at a
critical value, $U_{c}$, of the coupling constant (Hubbard $U$).
The SWF at the Fermi points $k_{F}$=$\pm{\pi }/2$ jumps, as $U$
increases, from a two-peak structure, the gap edges, to a
four-peak structure accompanied by a (non-vanishing) minimum of
the charge-gap. The two inner peaks of this structure show very
small dispersion (flat bands) away from the Fermi points, whereas
the outer peaks mark the edges of the Hubbard bands. No other
signatures of abrupt change are detected in the SWF. The two
regimes are physically realized in the angle-resolved
photoelectron spectra of $(TaSe_{4})_{2}I$, and the blue-bronze
$K_{0.3}MoO_{3}$, respectively.
\end{abstract}

\pacs{71.10.-w, 71.10.Fd, 71.10.Hf, 71.30.+h}

\maketitle

\section{ Introduction}

Quasi-one-dimensional (Q1D) systems have been the object of
intense experimental and theoretical activity over the last twenty
years. They show highly anisotropic properties, with a privileged
direction of enhanced charge transport. Their interest lies in the
hope that they can be good candidates for the physical realization
of non-Fermi liquid behavior. This interest, in low-D systems in
general, has expanded very rapidly in recent years partly due to
the technological development of low-D artificial structures and
nano-scale materials.

Above their Peirls temperature (or when doped away from half
filling), these Q1D systems are conductors and display Luttinger
liquid behavior\cite{haldane}, i.e., the absence of quasi-particle
excitations in the Fermi liquid sense (a quasi-particle peak at
the Fermi level) and the excitation, instead, of decoupled
collective modes of charge (holons) and spin (spinons) character,
a phenomenon usually known as charge-spin separation. The absence
of a Fermi edge has indeed been found in angle-resolved
photoelectron spectroscopy (ARPES) of
$(TaSe_{4})_{2}I$\cite{hwu,terra}, $K_{0.3}MoO_{3}$\cite{gweon},
and the organic conductor TTF-TCNQ
(tetrathiafulvalene-tetracyanoquinodimethane)\cite{zwick,sing}.
For a good review, see Ref 7. Clear experimental signatures of
spin-charge separation are, however, very scarce in Q1D
conductors, with the notable exception of TTF-TCNQ reported very
recently\cite{sing}. All these compounds in their metallic state
can be modeled by the low-energy physics of the doped 1D
single-band Hubbard Hamiltonian\cite{sing}. Alternatively, they
have been analyzed on the basis of the Luttinger model or the
Luther-Emery model (when a spin gap is expected). Some puzzles
still remain unsolved\cite{grioni}

Below their Peirls temperature, these Q1D compounds, as well as
many others like halogen-bridged transition-metal chains,
conjugated polymers, and organic charge-transfer salts are usually
insulating. In these systems, the competition between strong
on-site correlations and their kinetic energy gives rise to
significant localization of their itinerant electrons. This often
leads to the stabilization of non-metallic ground states with or
without charge-density waves (CDW). Thus $(TaSe_{4})_{2}I$ and the
blue bronze $K_{0.3}MoO_{3}$, for instance, are CDW
insulators\cite{grioni}, while the nearly ideal 1D CuO chains in
$SrCuO_{2}$ and $Sr_{2}CuO_{3}$ are responsible for the insulating
character of these charge-transfer insulators\cite{penc}.
Signatures of spin-charge separation have also been found in ARPES
of $SrCuO_{2}$\cite{kim} and in the dielectric response of
$Sr_{2}CuO_{3}$\cite{neudert}.

Most of these insulating systems can be conveniently described by
the Emery model\cite{emery}, which is a generalized Hubbard
Hamiltonian on a two-sublattice model, made of cations (say Cu
ions) and anions (say O ions), respectively. The on-site energy
levels and repulsions (${ \epsilon _{d}}$,$U_{dd}$) and (${
\epsilon _{p}}$,$U_{pp}$), for Cu $d$ orbitals and O $p$ orbitals,
are coupled by nearest-neighbor hopping of strength $t_{dp}$ and
Coulomb repulsions $V_{dp}$. Its 1D version reads

\begin{equation}
 H = \varepsilon _d \sum\limits_{is} {n_{dis}  + U_{dd}
\sum\limits_i {n_{di \uparrow } n_{di \downarrow }  + } }
\varepsilon _p \sum\limits_{js} {n_{pjs}+ U_{pp} \sum\limits_j
{n_{pj \uparrow } n_{pj \downarrow }} }
\end{equation}
\[
+t_{dp} \sum\limits_{ < ij > s} {(d_{is}^ +  p_{js}  + hc)}  +
V_{dp} \sum\limits_{ < ij > } {n_{di} n_{pj} }
\]where $ d_{is}^+$ creates an electron (hole) in a $d$ orbital at
site $i$ with spin $s$, $i$ running over all the Cu sites.
Similarly $p_{js}$ anihilates an electron (hole) in a $p$ orbital
at site $j$ and spin $s$, $j$ running over all the O sites. As
usual, $<ij>$ means summation over nearest-neighbors,
$n_{s}=c_{s}^ + c_{s}$  and $n_{i}=n_{i\uparrow} +
n_{i\downarrow}$ denotes the charge at the $ith$ site.

We may now simplify this hamiltonian. At first sight the Cu-O
repulsion seems essential since a large charge-transfer is
expected. However, $V_{dp}$ is usually much smaller than the
on-site repulsions and, furthermore, this term gives rise to new
physics only in the event of exciton formation. Hence, if we are
not especially interested in these processes, $V_{dp}$ can be
safely ignored. One is then left with a charge-transfer model
Hamiltonian which can still give a reasonable description of some
of these compounds\cite{zanen,meinders}, e.g., $SrCuO_{2}$ and
$Sr_{2}CuO_{3}$\cite{grioni}. If we are now willing to reduce the
number of parameters by putting (somewhat arbitrarily)
$U_{pp}=U_{dd}$, then the so-called ionic Hubbard model (IHM)
follows. It can be written simply as
\begin{equation}
H =  - t\sum\limits_{ < ij > s} {c_{is}^ +  c_{js}  + \frac{\Delta
}{2}}\sum\limits_{is} ( - 1)^i n_{is}  + U\sum\limits_i {n_{i
\uparrow } n_{i \downarrow } }
\end{equation}where $\Delta$ is the on-site energy difference
between even and odd sites, usually known as the charge-transfer
energy.

Although, strictly speaking, this Hamiltonian does not describe
accurately any specific system (since quite generally
$U_{pp}<U_{dd}$), it provides a simple, minimal, model where the
interplay among covalency ($t$), ionicity ($\Delta$) and
correlation ($U$) gives rise to a rich phase diagram within which
different 1D compounds can be placed. Originally proposed by
Nagaosa and Takimoto\cite{naga} as a model for ferroelectric
perovskites and later by Egami et al\cite{ega,ishi} to explain the
neutral-ionic transition in some organic crystals, this
Hamiltonian is ideal for studying the nature of quantum phase
transitions in 1D electron systems. On general grounds, one
expects a transition from an ionic, weakly-correlated band
insulator (BI) phase to a neutral, strongly-correlated Mott
insulator (MI) phase as $U$ increases. An important and
controverted issue is the nature of this transition as well as
whether two critical points rather than one separate both phases.
Depending on the method of calculation used, either
one\cite{resta,wilkens,kampf} or two\cite{fabrizio,taka,zhang,man}
critical points have been predicted, so that the controversy
cannot be considered as closed yet.

In this paper we stay outside this controversy and will rather
concentrate on the single-particle spectral-weight function (SWF)
$A(k,\omega)$ of the 1D IHM at half filling, which can be compared
with ARPES of several insulating materials. $A(k,\omega)$ will be
calculated using the cluster perturbation theory (CPT) approach of
Senechal et al \cite{sene}. This is briefly described in Sec II.
Sec III gives our results for $A(k,\omega)$ (a sudden change of
regime at a critical value of $U$) followed by a discussion
showing that $(TaSe_{4})_{2}I$ and $K_{0.3}MoO_{3}$, are good
examples of these two regimes. Finally, Sec IV closes the paper
with some concluding remarks.

\begin{figure}
\begin{center}
\includegraphics[scale=.5]{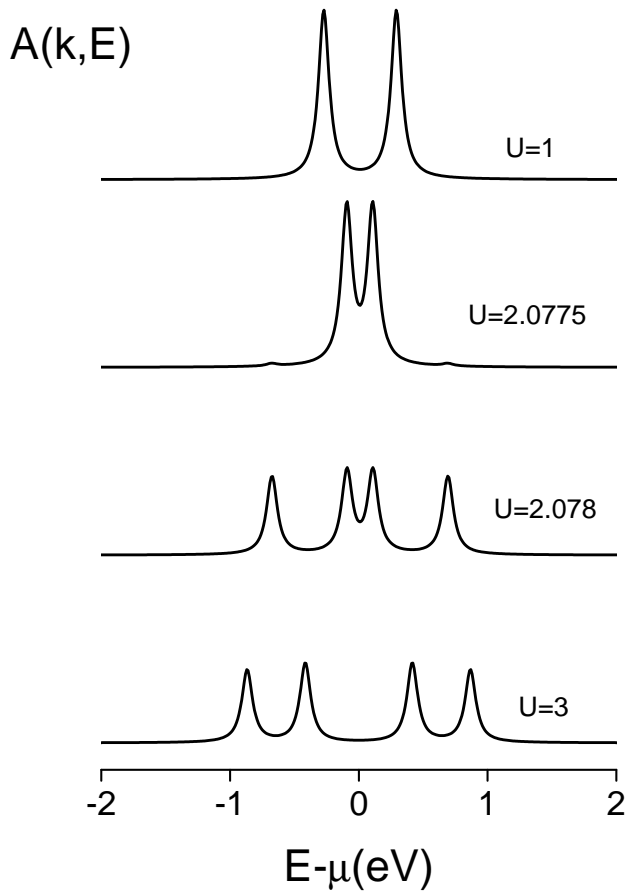}
\caption{Single-particle spectral-weight function $A(k,E)$ for the
half-filled ionic Hubbard model at the Fermi points $k=\pm \pi/2$.
From top to bottom, $U=1$, 2.0775 (just below $U_c$), 2.078 (just
above $U_c$), and 3. We have taken $t=-1$ and $\Delta=1$. All the
energies are given in eV.}
\end{center}
\end{figure}

\section{ Cluster perturbation theory (CPT)}

Since this method has been discussed at length in Ref 24, we
simply summarize it very briefly here. In CPT one divides the
lattice (here the chain) into a number of equal clusters. The
single-particle Green's function (GF) on these clusters is then
found by exact diagonalization with open boundary conditions. We
have made use of a variant of the Lanczos algorithm specifically
designed to calculate dynamic quantities\cite{dagotto}. The
approximation now consists in neglecting the intercluster
self-energy, so that the GF's of neighboring clusters are
connected by hopping terms only. Periodic boundary conditions are
then imposed on the whole chain, i.e., between the extreme
clusters. To be specific, let $mi$ denote the site $m$ of the
cluster $i$. The exact Green´s function $G_{mi,nj}$, of the whole
chain is given by the well-known Dyson's equation in matrix form
$(G_{0}^{-1}-\Sigma)G=I$, in terms of the non-interacting GF and
the exact $\Sigma$. In CPT this exact $\Sigma$ is approximated by
$\Sigma_{mi,nj}=\delta_{ij}\Sigma_{mn}^{C}$ where $\Sigma^{C}$ is
the cluster self-energy matrix. This approximation is applicable
to any lattice in any dimension. It can be understood as a
lowest-order contribution to a systematic perturbation expansion
in powers of the intercluster hopping\cite{sene,pair}. It turns
out, on the other hand, that CPT is a limiting case of a more
general variational cluster approach\cite{pott}.

In this paper we concentrate on the SWF $A(k,\omega)$, given as
usual by
\begin{equation}
A(k,\omega ) = -\frac{1}{\pi }{\mathop{\rm Im}\nolimits}
G(k,\omega + i\eta )
\end{equation}where $G(k,\omega)$ is the Fourier transform (FT) of
the single-particle retarded GF and $\eta$ a small positive
number. This FT must be calculated with some care since
$G_{mn}(i-j)$ is periodic in $ij$ but not in $mn$ due to the open
boundary conditions used in the clusters. The correct formula
is\cite{sene,pair}
\begin{equation}
G(k,\omega ) = \frac{1}{N}\sum\limits_{mn} {e^{ - ik(m - n)}
G_{mn} } (Nk,\omega )
\end{equation}where $N$ is the number of sites in a cluster and
$G_{mn}(k)$ the FT of $G_{mn}(i-j)$.

\begin{figure}[th]
\begin{center}
\includegraphics[scale=.5]{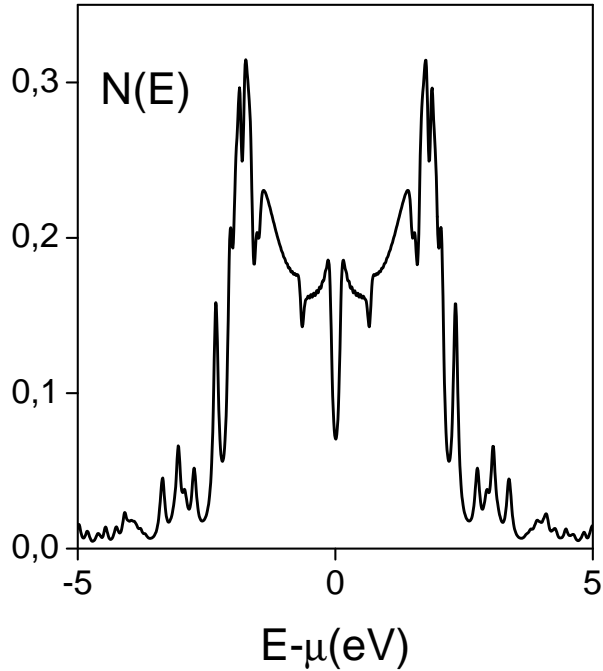}
\caption{Density of states for the same model of Fig1 with $U=
2.0775$, eV just below $U_{c}$.}
\end{center}
\end{figure}

\section{ The spectral weight function of the ionic Hubbard model at half filling}.

We consider a chain of ninety-six sites of two kinds with levels
at $\pm\Delta/2$, at half filling, and take $t=-1eV$ and
$\Delta=1eV$. Clusters of eight sites have been adopted after
checking that increasing the cluster size up to twelve sites does
not change much the results for the whole chain. Fig 1 shows
$A(k,E)$ at the Fermi points $k_{F}=\pm\pi/2$ for increasing $U$.
A broadening $\eta=0.05$ eV has been given to the otherwise delta
functions. Two regimes are observed in the low-energy region: For
small $U$, a two-peak structure is seen close to the chemical
potential $\mu$ (zero reference-energy). This structure persists,
while the two peaks approach each other, up to $U=2.0775$eV.
Abruptly, at $U=U_{c}=2.078$ eV a four-peak structure appears,
with the peak heights being roughly half those of the previous
two-peak structure. It seems, therefore, that each of the latter
peaks splits, developing a new peak at higher energy and thus
leading to the new four-peak structure. As $U$ further increases,
the inner peaks, closer to the chemical potential, start to
separate from each other quickly approaching the outer peaks which
in turn move apart more slowly. Finally, for very large $U$, much
larger than $t$ and $\Delta$, the SWF tends asymptotically to a
Mott-Hubbard situation without any signature of abrupt change. The
single-particle (charge) gap, after passing through a minimum at
$U_{c}$, increases again. It never vanishes, in agreement with
recent findings\cite{kampf,man}.
\begin{figure}
\begin{center}
\includegraphics[scale=.5]{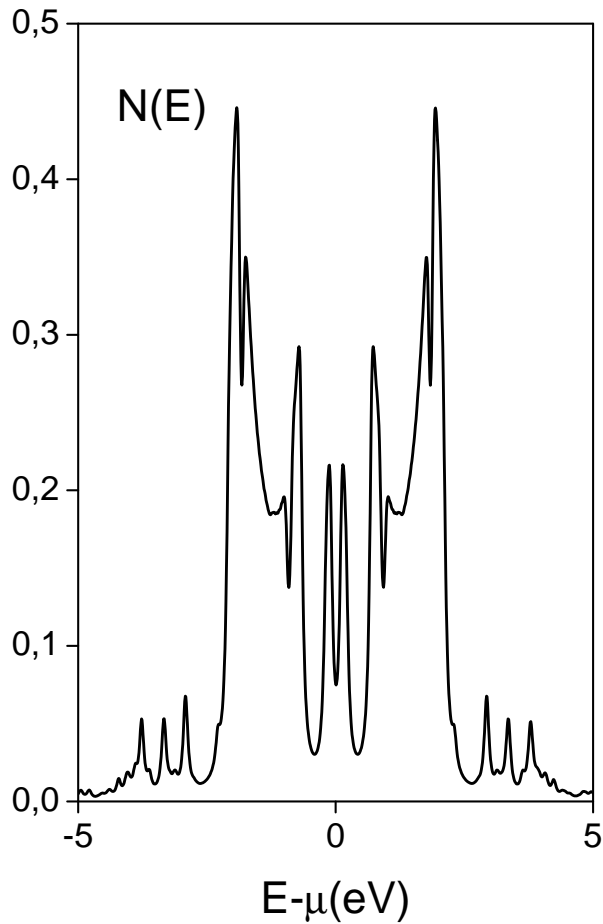}
\caption{The same as Fig 2 but for $U=2.0780$, eV just above
$U_{c}$}
\end{center}
\end{figure}

A somewhat different, but complementary perspective is afforded by
the density of states (DOS), $N(E)=(1/M)\sum_{k}A(k,E)$, where $M$
is the number of $k's$. Fig 2 shows the DOS for $U=2.0775eV$, just
below $U_{c}$. It is reminiscent of the non-interacting DOS (a
brunch-cut surrounded by two singularities and cut by a gap at the
chemical potential). This region $U<U_{c}$, is the band insulating
regime, the DOS being similar to that for $U=0$, but with a
continuously decreasing gap.  Fig 3, on the other hand, displays
the DOS for $U=2.078eV$, just above $U_{c}$, in correspondence
with the third panel (from top to bottom) of Fig 1. The two outer
peaks in this latter figure mark the top (bottom) of the occupied
(empty) Hubbard band. The two inner peaks stand isolated near the
chemical potential resembling two "impurity-like", non-dispersive
peaks within the (wider) gap between the Hubbard  bands. All
through the region $U>U_{c}$, the DOS is quite different from that
at $U=0$.

\begin{figure}
\begin{center}
\includegraphics[scale=.5]{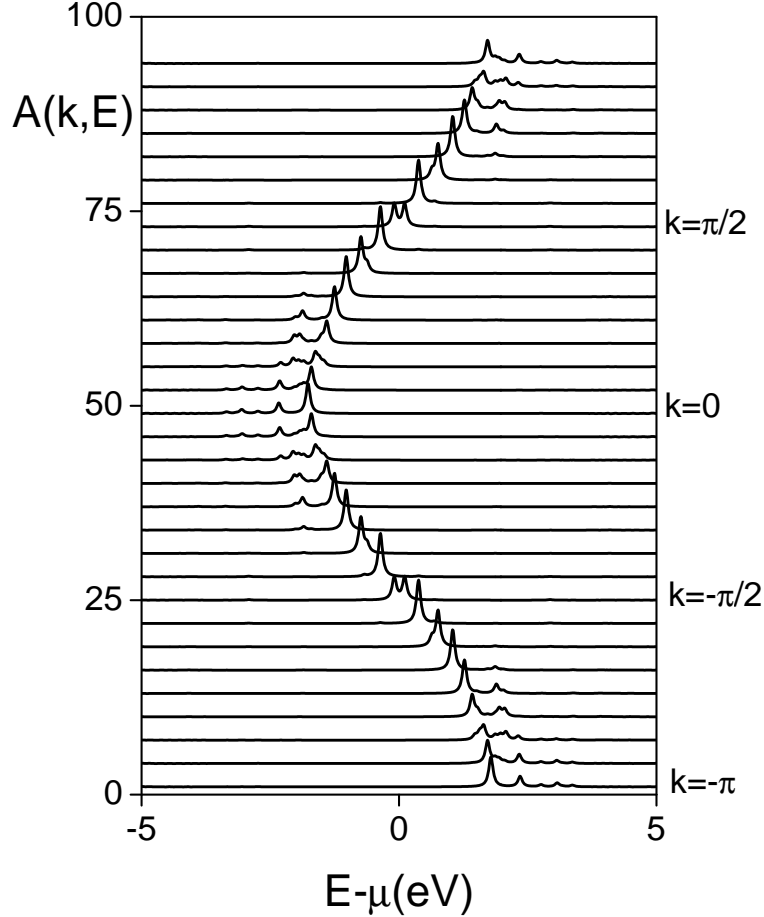}
\caption{Spectral-weight function $A(k,E)$ for the same situation
of Fig 2. An offset has been given to the plots for different k's
along the large Brillouin zone $(-\pi\leq k<\pi)$ in order to
avoid superposition. The figures along the left vertical axis
count the number of k's starting from $-\pi$. On the right
vertical axis, some especial k's are indicated}
\end{center}
\end{figure}

To check what changes have taken place at $k\neq\pm\pi/2$, Figs 4
and 5 display A(k,E) for the same $U$ values as the above DOS
along the large Brillouin Zone (BZ), $-\pi\leq k\leq\pi$ (in the
extended zone scheme). The portion $\mid k\mid
>\pi/2$ can be mapped, if one wishes, onto the small BZ giving
the empty bands for $\mid k\mid <\pi/2$ above the occupied bands.
An offset has been provided to the different plots to avoid
superposition. The figures along the vertical axis number the k's
starting from $k=-\pi$ (only a selected set of thirty-two k's have
been shown for clarity). Fig 4 shows a cosine-like band cut in two
by a gap at the Fermi points (cfr. with the second panel in Fig
1). Two shadow bands covering only part of the BZ are clearly
visible around $k=0$ (occupied) and $k=\pi$ (empty). They give
rise to the side-peaks to the left and right of the main body of
the DOS of Fig 2. Fig 5 should now be confronted with the DOS of
Fig 3. The cosine-like band is now cut by a wider gap delimited by
the outer peaks of Fig 1, third panel. The inner peaks are
continued by two almost non-dispersive, flat bands covering only a
small portion of the BZ around $k_{F}$ are clearly visible. The
shadow bands around $k=0$ and $k=\pi$ have now almost disappeared
along with the side-peaks, rather weak in Fig 3.

\begin{figure}
\begin{center}
\includegraphics[scale=.5]{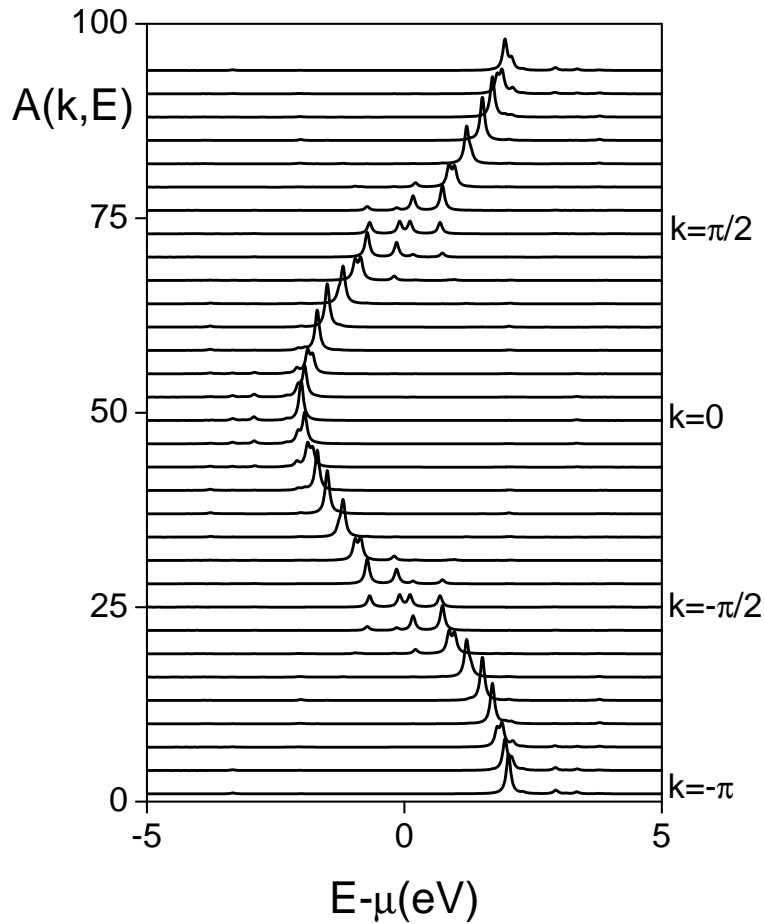}
\caption{Same as Fig 4, but for $U=2.0780$, eV just above $U_{c}$
}
\end{center}
\end{figure}
It is difficult to extract the character of the peaks in a
calculation based on a Lanczos algorithm, and much more so to
trace the origin of the change of regime just described. However,
these regimes must have some physical reality since we observe
that both are found in ARPES of different Q1D insulating
materials. Since most of these materials have occupied bands with
roughly $1 eV$ of bandwidth, we have accordingly taken $t=-0.5eV$
and $\Delta=1eV$. The critical $U$ separating both regimes is now
$U_{c}=2eV$. Let us take, just for illustrative purpose,
$(TaSe_{4})_{2}I$, and $K_{0.3}MoO_{3}$, Fig 6 shows $A(k,E)$
between $k=0$ and $\pi/2$ for $U=0.5eV<U_{c}$, a case of the first
regime, with a single occupied peak at $k=\pi/2$. This one-band
structure resembles that of $(TaSe_{4})_{2}I$ in the direction
parallel to the chain (compare with Fig 2a of Ref 3). Likewise.
Fig 7 shows the same information as Fig 6, but for
$U=2.1eV>U_{c}$. We now see a two-band structure in the
neighborhood of $k=\pi/2$. This resembles the band structure of
$K_{0.3}MoO_{3}$ (compare with Fig 15 of Ref 4).
\begin{figure}
\begin{center}
\includegraphics[scale=.5]{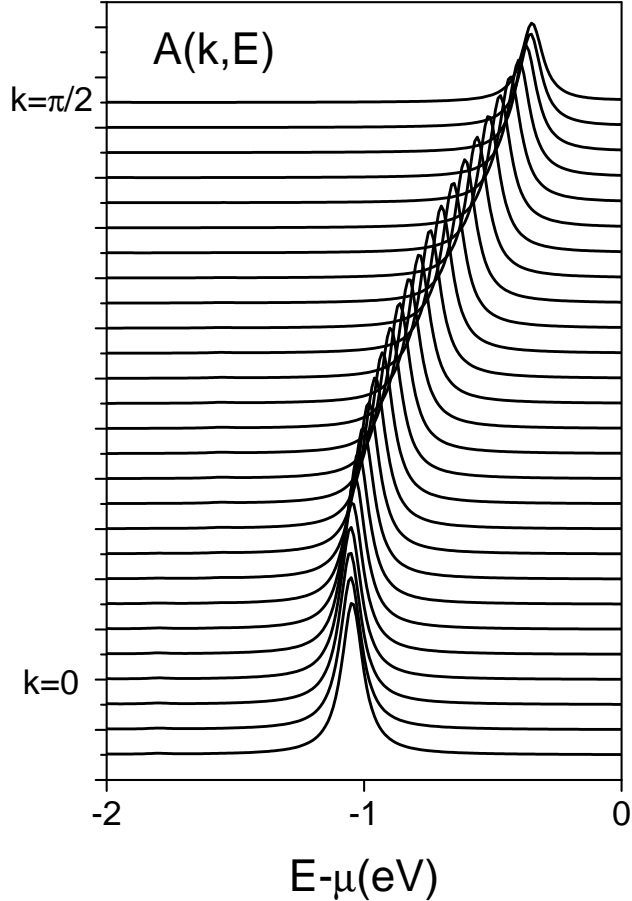}
\caption{Spectral-weight function for (in eV) $t=-0.5$, $\Delta=1$
and $U=0.5$}
\end{center}
\end{figure}
\begin{figure}
\begin{center}
\includegraphics[scale=.5]{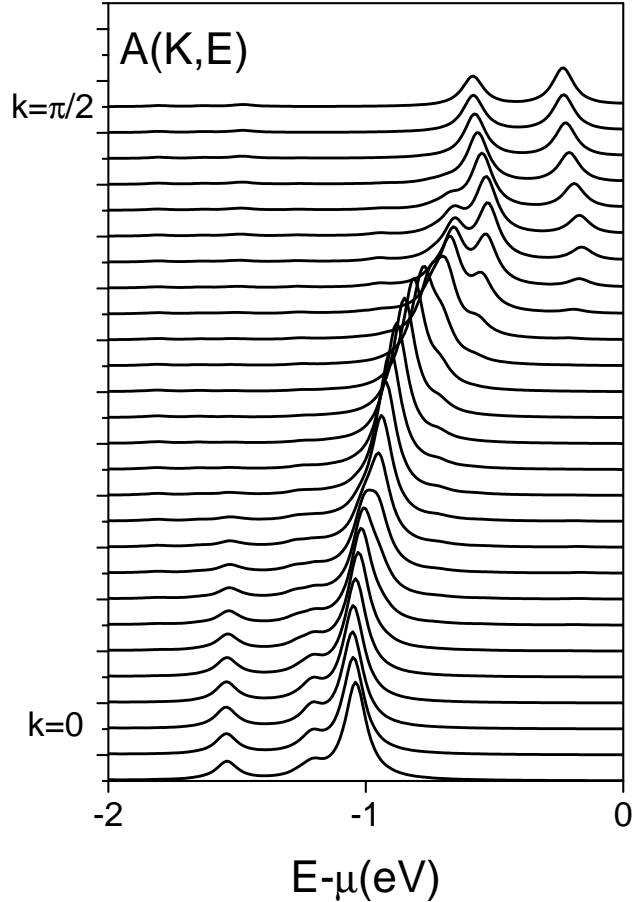}
\caption{The same as Fig 6, but for $U=2.1$ eV}
\end{center}
\end{figure}

\section{ Conclusions}
Using the cluster perturbation theory approach of Senechal et
al\cite{sene,pair}, we have calculated the single-particle
spectral-weight function $A(k\omega)$ of the ionic 1D Hubbard
model at half filling. A change of regime is found at a critical
value of $U$, $U_{c}(t,\Delta)$, which depends on both the hopping
amplitude $t$ and the on-site energy-difference, $\Delta$, between
even and odd sites. As $U$ increases, $A(k\omega)$ jumps from a
two-peak structure to a four-peak one at the Fermi points
$k_{F}=\pm\pi/2$. As one moves away from $k_{F}$, one finds two
semiconducting bands, for $U<U_{c}$, separated by a gap which
decreases from its initial value $\Delta$ (at $U=0$) down to a
small, but non-vanishing, value at $U_{c}$. This gap is always
delimited by the two-peak structure at $k_{F}$. For $U>U_{c}$,
instead, two flat, almost-non-dispersive bands appear around the
Fermi points as the continuation of the inner peaks in the
four-peak structure. They fade away very soon. The gap between
them now increases from its minimum value at $U_{c}$. The outer
peaks mark the top (bottom) of the lower (upper) Hubbard-like
bands. As $U$ increases further, the flat bands approach the
Hubbard bands, finally merging into them. Asymptotically, a
Mott-Hubbard situation is approached in a continuous way, without
any signature of abrupt change in $A(k,\omega)$ in this region of
large $U$.

Alternatively we can say that, for $U<U_{c}$, the two-band
structure is continuously connected to the band insulator shape at
$U=0$. This is the band insulator regime. For $U>U_{c}$, two new
flat bands appear and push the two wider bands of the first regime
further apart. This is the second regime, more Hubbard-like, which
goes to the Mott-Hubbard regime at large $U$. Different Q1D
materials have band structures which can be classified as lying in
either the first or the second regime.

We acknowledge the financial support of the Spanish DGICYT through
Project BFM 2002-01594

\end{document}